\documentclass{elsart}

\usepackage{amssymb}
\usepackage{graphicx}

\newcommand\0{\over } \newcommand\2{{1\over2}}

\newcommand{\bea}{\begin{eqnarray}}
\newcommand{\eea}{\end{eqnarray}}
\newcommand{\be}{\begin{equation}}
\newcommand{\ee}{\end{equation}}

\begin{document}
\begin{flushright}
~\vspace{-1.25cm}\\
{\small\sf
TUW-02-24\\[-2pt]
YITP-SB-02-62
}
\end{flushright}
\vspace{0.5cm}
\begin{frontmatter}
\title{Comment on ``One Loop Renormalization of Soliton Quantum Mass
Corrections in 1+1 Dimensional Scalar Field Theory Models''\\
{\large\bf (Phys. Lett. B542 (2002) 282 [hep-th/0206047])}
}

\author[TUW]{A. Rebhan}, \author[YITP]{P. van Nieuwenhuizen}, \author[TUW]{R. Wimmer}
\address[TUW]{Institut f\"ur Theoretische Physik,
         Technische Universit\"at Wien,\\
         Wiedner Hauptstra\ss e 8-10/136,
         A-1040 Vienna, Austria
}%
\address[YITP]{C.~N.~Yang Institute for Theoretical Physics, \\
  SUNY at Stony Brook, Stony Brook, NY 11794-3840, USA}

\begin{abstract}
We refute the claim that previous works on the one-loop quantum
mass of solitons had incorrectly dropped a surface term from
a partial integration. Rather, the paper quoted in the title
contains a fallacious derivation 
with two compensating errors. We also remark that the 
$\phi^2\cos^2\ln(\phi^2)$ model considered in that paper
does not have solitons at the quantum level because
at two-loop order the degeneracy of the vacua is lifted.
This may be remedied, however, by a supersymmetric extension.
\end{abstract}

\end{frontmatter}


The issue of quantum corrections to solitons, both in bosonic
and in supersymmetric theories, has in the last few years
received extensive examination and a number of subtleties
have been revealed and clarified. 
In a recent publication in Physics Letters B,
Ref.~\cite{Flores-Hidalgo:2002at},
a formula for the quantum mass of (1+1)-dimensional solitons at
one-loop order has been put forward, which includes a surface term from
a partial integration that allegedly had not been considered in other
(and therefore incorrect)
treatments. 
In this Comment we wish to make it clear (a) that
the previous works quoted in Ref.~\cite{Flores-Hidalgo:2002at}
did include this surface term where appropriate, (b) that
this surface term arises only in certain regularization prescriptions
for sums of zero-point energies, and
(c) that the actual derivation presented in Ref.~\cite{Flores-Hidalgo:2002at}
is fallacious and leads to the correct result 
only because of two compensating errors, which unfortunately
can also be found in the widely used textbook \cite{Raj:Sol}, as two of us have
pointed out already
in a footnote in Ref.~\cite{Rebhan:1997iv}.

Whereas in the case of solitons in scalar (1+1)-dimensional 
field theories the correct one-loop results are known from
the classic papers of Ref.~\cite{Dashen:1974cj,Dashen:1975hd},
the situation is much more involved and subtle in the
corresponding supersymmetric (SUSY) models. In fact,
contradictory results \cite{D'Adda:1978mu,Schonfeld:1979hg,Rouhani:1981id,Kaul:1983yt,Imbimbo:1984nq,Chatterjee:1984xh,Yamagishi:1984zv,Uchiyama:1986gf,Uchiyama:1986ki,Boya:1988zh,Casahorran:1989vd}
have dominated the early literature on
one-loop corrections to (1+1)-dimen\-sional SUSY solitons,
which in part were due to a surprising sensitivity
to the regularization method even when the renormalization
conditions have been fully fixed \cite{Rebhan:1997iv}.
This finding has triggered a number of investigations, and
the subtleties of the various methods that can be employed
to calculate the one-loop corrections of (1+1)-dimensional
SUSY solitons have been sorted out in every detail
only rather recently \cite{Nastase:1998sy,Graham:1998qq,Shifman:1998zy,Litvintsev:2000is,Goldhaber:2000ab,Goldhaber:2001rp,Wimmer:2001yn,Bordag:2002dg,Goldhaber:2002mx,Rebhan:2002uk,Rebhan:2002yw}.
It is therefore unfortunate that a new paper on
the simpler (1+1)-dimensional scalar models with solitons,
where some of these issues already arise, ignores these
subtleties and misinterprets the literature that
has in fact resolved them. 
We therefore deem it necessary
to put things right and 
thus prevent the possible spread of
new confusion about the status of the various methods
to calculate one-loop corrections to soliton masses.

In the notation of Ref.~\cite{Flores-Hidalgo:2002at}, the
one-loop contribution to the quantum mass of a soliton arises
from the difference in zero-point energies in the presence of
the soliton and in the vacuum,
\be\label{Msum}
\Delta M_{\rm bare}=\2\sum_i \omega_i+\2\sum_q \omega(q)
-\2\sum_k \omega^0(k),
\ee
where $i$ and $q$ labels the eigenfrequencies of 
the discrete and continuum modes
in the presence of a soliton, and $k$ those of the vacuum
modes. The energy-momentum relation is formally the same
for the soliton and the vacuum sector,
$\omega(q)=\sqrt{q^2+m^2}$ and $\omega^0(k)=\sqrt{k^2+m^2}$,
but in the soliton background there is an additional
phase shift $\delta(q)$ which gives rise to different
densities of states.

Because the sums over zero-point energies are quadratically divergent,
they, as well as
their difference, which still contains a logarithmic divergence,
require careful regularization. In mode regularization \cite{Dashen:1974cj}
the system is put into a finite box, and if one considers
an equal (and finite) number of modes in the two sectors,
one can, as done in Ref.~\cite{Flores-Hidalgo:2002at},
consider individual differences of zero-point energies
with given (e.g. periodic) boundary conditions, leading to
\be\label{dq}
\sqrt{q_n^2+m^2}=\sqrt{k_n^2+m^2}-{k_n \delta(k_n)\0L\sqrt{k_n^2+m^2}}
+O(L^{-2}).
\ee

Inserting this in (\ref{Msum}) and using the free density
of states $L/(2\pi)$, Ref.~\cite{Flores-Hidalgo:2002at}
states that an integration by part gives
\be\label{FHres}
\Delta M_{\rm bare}=\2\sum_i \omega_i
-{\omega_k \delta(k)\04\pi}\bigg|_{-\infty}^\infty
+\2\int_{-\infty}^\infty{dk\02\pi}\omega(k){d\0dk}\delta(k),
\ee
and that the resulting surface term
was not considered in other treatments, for example
\cite{Graham:1998kz}, \cite{Goldhaber:2000ab}, \cite{Rebhan:1997iv}.

\begin{figure}[b]
\centerline{\includegraphics[bb=72 320 525 480,width=9cm]{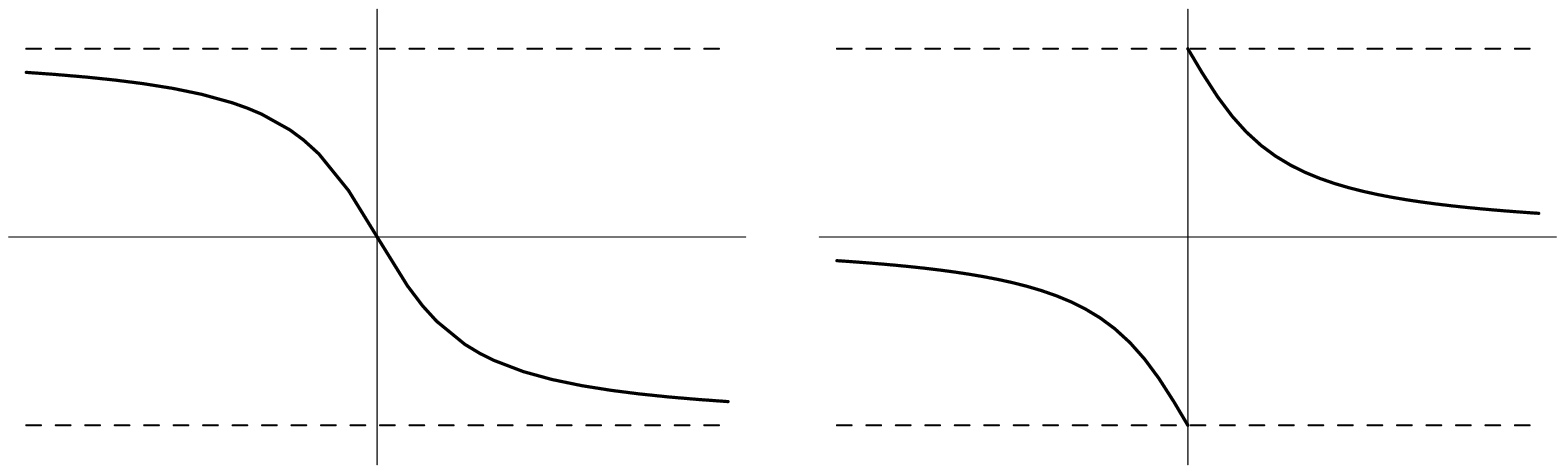}}
\centerline{\hfil\quad (a) \hfil\hfil (b) \hfil}
\caption{Schematically, two possible choices for the phase shift
function $\delta(k)$: (a) continuous, with $\delta(+\infty)-\delta(-\infty)=
-2\pi \mathcal N$, $\mathcal N\ge1$; (b) discontinuous
with $\delta(0+)-\delta(0-)=2\pi\mathcal N$.
Ref.~\cite{Flores-Hidalgo:2002at} implicitly assumed both, a continuous
behaviour and $\delta(k\to\pm\infty)\to0$, which cannot be had
at the same time.
\label{fig}}
\end{figure}

While this latter part of the statement is correct (if empty, see below)
as concerns
Ref.~\cite{Graham:1998kz}, it is incorrect for the other two
references given at that point. In particular, Ref.~\cite{Rebhan:1997iv}
discusses this surface term at length and moreover shows that the above
derivation, which can also 
be found in the textbook of Rajaraman \cite{Raj:Sol},
is not correct. The above derivation treats $\delta(k)$ as a
continuous function, but then it cannot vanish for $k\to\pm\infty$
because a continuous phase shift function has
$\delta(+\infty)-\delta(-\infty)=-2\pi \mathcal N$, where $\mathcal N\ge1$
is the number of discrete modes, and the surface term
would be divergent.
In this case one has to compare $k_n$ and $q_n$ with shifted
mode numbers such that at large $k$ the effective phase shift
in (\ref{dq}) tends to zero
\cite{Rebhan:1997iv}.
An alternative is clearly to adopt a
discontinuous $\delta(k)$ such that $\delta(k\to\pm\infty)\to 0$,
and one may choose to put the discontinuity at $k=0$
(see Fig.~\ref{fig}). In this approach
there is a further surface term contributing to (\ref{FHres}), namely
$${\omega_k \delta(k)\04\pi}\bigg|_{0-}^{0+}=\2m \mathcal N.$$
With a finite number of modes in both sectors, and the requirement
of their equality, this extra contribution is cancelled because 
the vacuum sector has $\mathcal N$ more continuous modes
than the soliton sector, and these need to be subtracted explicitly,
as discussed in detail in Ref.~\cite{Nastase:1998sy,Goldhaber:2000ab}.
Clearly, these considerations require regularization (a finite
mode number cutoff which is taken to be the same in
the two sectors), but Ref.~\cite{Flores-Hidalgo:2002at} skips
the stage where the regulator is still in place.

Now it is true that in other treatments the above surface terms
are not considered. The reason is simply that they arise only
in mode regularization in a finite volume.
If one refrains from introducing a finite volume
to make the spectrum discrete, one naturally {\em starts} from
considering the difference of the spectral densities which
is just given by $\delta'(k)$ and uses this to integrate
over the continuous spectrum, i.e. (\ref{FHres}) {\em without}
the surface term. This only makes sense if the $k$-integration
is regularized, and explicit calculation for example in
dimensional regularization \cite{Parnachev:2000fz,Rebhan:2002uk}
does give the correct result. 

Simple energy/momentum cutoff 
regularization in such a continuum formulation, on the
other hand, is more dangerous, as has been shown in Ref.~\cite{Rebhan:1997iv},
and it has led to incorrect results in the early literature
on SUSY solitons.
It can, however, be repaired by introducing a sharp cutoff on
the phase shift function as a limit
of smooth ones \cite{Litvintsev:2000is}, which produces in fact an
additional term equivalent to the one in (\ref{FHres}). However,
it is not correct that this additional term is the same
as the surface term encountered in mode number regularization,
as suggested by the author of Ref.~\cite{Flores-Hidalgo:2002at} who
implies that the error of a naive momentum cutoff was
simply the omission of this surface term.
As we have seen, for a phase shift function
with $\delta(k\to\pm\infty)\to 0$,
the surface terms that Ref.~\cite{Flores-Hidalgo:2002at}
should have included involve
an additional contribution at low momentum that is not generated by
smoothing out the high momentum cutoff. 
And in momentum cutoff regularization
these extra
terms are not cancelled as in mode number regularization,
because the former by definition
does not require equality of the number of modes but instead equality
of the momentum cutoff in the soliton and vacuum sector.
The problem of naive momentum cutoff regularization is therefore
exclusively that of a sharp UV cutoff. As an aside, 
we have recently demonstrated \cite{Rebhan:2002uk} that
a smooth UV cutoff on the phase shift function
is in fact required to obtain finite
results in higher-dimensional kink domain walls; only in
1+1 dimensions a sharp UV cutoff presents a real pitfall
in that it leads to finite (but incorrect) results.

We would like to emphasize once more the importance of regularization
in the problem of quantum corrections to soliton masses. In
order to understand and resolve the subtleties arising in
this case, it is mandatory to avoid manipulation of unregulated
and thus ill-defined quantities. 
This is particularly necessary in the supersymmetric case:
In Refs.~\cite{Rebhan:1997iv,Nastase:1998sy} two of us have
encountered the problem that nonzero corrections to the
quantum mass of SUSY solitons are seemingly not matched by those in
the central charge operator as required
by BPS saturation \cite{Witten:1978mh}. In Ref.~\cite{Nastase:1998sy}
we conjectured the existence of an anomaly, and this was
subsequently established by Shifman et al.~\cite{Shifman:1998zy} in the form of
an anomalous additional term in the central charge operator.
On the other hand, in
Ref.~\cite{Graham:1998qq} Graham and Jaffe have put forward a proof that
the central charge and the mass of SUSY solitons receive
the same corrections, apparently without the need for an
anomalous contribution. However, Ref.~\cite{Graham:1998qq} 
arrived at this conclusion by formal manipulation of unregularized
quantities, invoking (dimensional) regularization only
at a later stage.
In Ref.~\cite{Rebhan:2002yw} we have
recently resolved this apparent contradiction
and demonstrated how the anomalous contribution to
the central charge of the SUSY kink arises from a careful
implementation of dimensional regularization.

Finally, we would like to remark that
Ref.~\cite{Flores-Hidalgo:2002at} also
contains an interesting generalization
to non-reflectionless potentials and the associated field theory
models in 1+1 dimensions. As one particular example,
a $V=\phi^2\cos^2\ln(\phi^2)$ model is considered which has
been introduced in Ref.~\cite{Flores-Hidalgo:2002ma}. Such a model has
infinitely many degenerate vacua, accumulating about $\phi=0$
which is a non-analytic point.
At one-loop order, the degeneracy of these vacua at $\phi\not=0$ is
not lifted because all minima have equal curvature $V''$, and only
this enters in the one-loop effective potential. However,
at two-loop order the quantum corrections are proportional
to $(V''')^2$, which is larger for the minima with smaller value
of $\phi$ (and diverges for $\phi\to0$). As a consequence, at the
quantum level neighboring minima are no longer degenerate and
the corresponding solitons do no longer exist. It does therefore,
unfortunately, make no sense to calculate quantum corrections
to the solitons as these only exist in the (semi-)classical theory.
However, a possibility
for keeping solitons at the quantum level in such a model
might be to consider
its supersymmetric extension as in Ref.~\cite{Kaul:1985yy}, where
a similar problem has been encountered in a simpler
model and at one-loop level. We intend to study this question further
in a separate work.

\end{document}